\documentclass[aps,prd,reprint,nofootinbib]{revtex4-2}

\usepackage{amsmath,amssymb,bm}
\usepackage{graphicx}
\usepackage{xcolor}
\usepackage[colorlinks=true,linkcolor=blue,citecolor=blue,urlcolor=blue]{hyperref}
\usepackage{soul}

\newcommand{\nn}{\nonumber\\}
\newcommand{\lb}{\left(}
\newcommand{\rb}{\right)}
\newcommand{\para}{\parallel}

\begin{document}

\title{Relativistic BDNK MHD Evolution in a Boost-Invariant Medium and Its Impact on Dilepton Production}

\author{Ankit Kumar Panda}
\email{ankitkumarpanda932@gmail.com}
\affiliation{Key Laboratory of Quark and Lepton Physics (MOE) \& Institute of Particle Physics, Central China Normal University, Wuhan 430079, China}

\author{Rajesh Biswas}
\email{rajesh.biswas@pwr.edu.pl}
\affiliation{Institute of Theoretical Physics, Wroc\l{}aw University of Science and Technology, 50-370 Wroc\l{}aw, Poland}



\begin{abstract}
In this work, we explore a Bemfica--Disconzi--Noronha--Kovtun (BDNK)-type formulation of relativistic magnetohydrodynamics, providing a causal and stable first-order description of dissipative fluids. We derive coupled evolution equations for the temperature and magnetic field in a boost-invariant Bjorken background, restricting to $(0+1)$D dynamics while retaining all relevant first-order gradients. By varying the transport coefficients, we disentangle the interplay and mutual backreaction between the thermal and electromagnetic sectors.
We find that, for comparable transport coefficients, the magnetic field responds more strongly to changes in the temperature evolution, while its feedback on the temperature remains subleading. We further analyze the number density evolution, which is sensitive to both temperature gradients and magnetic-field dynamics.
We also investigate implications for dilepton production, where the magnetic field modifies the emission rate via the relaxation time in a kinetic-theory framework. The coupled evolution leads to a suppression of the low-mass dilepton spectrum, primarily driven by enhanced cooling in the presence of positive coupling between temperature gradients and magnetic-field evolution, as compared to scenarios without such feedback.

\end{abstract}

\maketitle


\section{Introduction}

Relativistic heavy-ion collisions provide a unique laboratory to study strongly interacting matter under extreme conditions of temperature and density, where a deconfined phase of quarks and gluons—the quark--gluon plasma (QGP)—is formed. Experimental programs at the Relativistic Heavy Ion Collider (RHIC) and the Large Hadron Collider (LHC) have established that the QGP behaves as a nearly perfect fluid, exhibiting strong collective flow and remarkably small viscosity~\cite{Romatschke:2007mq,Song:2010mg,Heinz:2013th,Busza:2018rrf}.
Relativistic hydrodynamics has thus emerged as a successful framework for describing the spacetime evolution of the QGP~\cite{Song:2007ux,Qiu:2011hf,Niemi:2011ix}. As an effective theory of long-wavelength dynamics, it provides a quantitative link between the initial state and final-state observables such as particle spectra and flow coefficients. However, the relativistic formulation of first-order dissipative hydrodynamics, developed by Landau--Lifshitz~\cite{LandauLifshitzFluid} and Eckart~\cite{Eckart}, suffers from fundamental issues, including acausality and instabilities associated with superluminal signal propagation~\cite{Hiscock:1983zz,Hiscock:1985zz,Hiscock:1987zz}.
These limitations led to the development of second-order theories, most notably M\"uller--Israel--Stewart (MIS) hydrodynamics~\cite{Muller:1967zza,Israel:1976tn,Israel:1979wp}, along with modern refinements~\cite{Muronga:2001zk,Denicol:2012cn,Baier:2007ix,Bhattacharyya:2007vjd,Jaiswal:2013vta,Grozdanov:2015kqa}. These approaches restore causality and stability by introducing additional dynamical degrees of freedom and associated relaxation times~\cite{Denicol:2008ha,Pu:2009fj,Brito:2020nou,Bemfica:2020xym,Roy:2025nay}.
More recently, a first-order formulation due to Bemfica, Disconzi, Noronha, and Kovtun (BDNK)~\cite{Bemfica:2017wps,Bemfica:2019knx,Bemfica:2020zjp,Kovtun:2019hdm} has been shown to be both stable and causal without invoking higher-order gradients. This is achieved by consistently incorporating temporal and spatial gradients at first order within a generalized hydrodynamic frame. There have also been several important developments in this direction, extending, refining and validating the framework in various contexts~\cite{Biswas:2022cla,Biswas:2022hiv,Pandya:2021ief,Rocha:2022ind,Hoult:2020eho,Hoult:2021gnb,Abboud:2023hos,Hoult:2024cyx,Weickgenannt:2023btk,Jain:2023obu,Armas:2022wvb,Hoult:2024qph}.

In non-central heavy-ion collisions, the formation of the quark--gluon plasma (QGP) is accompanied by the generation of extremely strong electromagnetic fields, primarily arising from the relativistic motion of charged protons within the colliding nuclei~\cite{Skokov:2009qp,Bzdak:2011yy,Deng:2012pc,Gursoy:2014aka,Panda:2024ccj,Dash:2023kvr}. At early times, the magnitude of the magnetic field can be comparable to intrinsic QCD scales, making it capable of significantly influencing the dynamical evolution of the plasma.
A natural framework to describe such systems is relativistic magnetohydrodynamics (MHD)~\cite{Roy:2015kma,Pu:2016ayh,Mayer:2024dze,Hongo:2013cqa,Inghirami:2016iru,Inghirami:2019mkc,Siddique:2019gqh,Pu:2016bxy,Roy:2017yvg,Nakamura:2022wqr,Mayer:2024kkv}, which provides a self-consistent macroscopic description of the coupled evolution of matter and electromagnetic fields. Further theoretical developments in relativistic MHD can be found in Refs.~\cite{Hernandez:2017mch,Grozdanov:2016tdf,Grozdanov:2018fic,Grozdanov:2017kyl,Panda:2020zhr,Panda:2021pvq,Denicol:2018rbw,Denicol:2019iyh,Kushwah:2025jsb,Singh:2024leo,Tiwari:2025xrs,Biswas:2020rps,Cordeiro:2023ljz,Hattori:2022hyo}. More recently, the BDNK framework has been extended to incorporate magnetic fields at first order~\cite{Armas:2022wvb,Hoult:2024qph}, enabling a causal and stable treatment of electromagnetic backreaction within a first-order hydrodynamic theory.
Such a framework is particularly important from a phenomenological perspective, as electromagnetic fields are expected to leave measurable imprints on experimental observables~\cite{Harris:2023tti,Panda:2023akn,Panda:2025lmd,Parida:2025ddt,Bagchi:2026aqu,Panda:2026kko}. Among these, dileptons~\cite{McLerran:1984ay,Rapp:2014hha,Shuryak:1978ij} serve as especially valuable probes, since they interact only electromagnetically and thus carry largely undistorted information from the entire spacetime evolution of the medium. In this context, magnetic fields can modify microscopic interaction rates---for instance, through changes in effective relaxation times~\cite{Panda:2020zhr,Dwibedi:2025xho}---thereby influencing both dilepton yields and their anisotropic flow.

Thus in the current work, we explore a self-consistent causal first-order BDNK--MHD framework to study the coupled evolution of temperature, particle number density, and magnetic fields. To achieve analytical simplicity and numerical control, we consider a boost-invariant Bjorken expansion in Milne coordinates and solve the resulting nonlinear system of ordinary differential equations. We then investigate how this coupled evolution influences dilepton production, with particular emphasis on the role of electromagnetic fields and their backreaction on the medium. This approach enables us to assess the phenomenological implications of causal first-order MHD dynamics in relativistic heavy-ion collisions.

The paper is organized as follows. In Sec.~\eqref{1}, we introduce the BDNK formulation of relativistic MHD and present the first-order gradient expansion of the energy-momentum tensor, the electromagnetic field strength, and the number current. We then derive the corresponding reduced set of evolution equations within a boost-invariant $(0+1)$D Bjorken framework in the latter part of this section. 
In Sec.~\eqref{2}, we systematically analyze the coupled evolution of temperature, magnetic field, and number density by varying the relevant transport coefficients, thereby isolating the effects of temperature-to-magnetic field and magnetic field-to-temperature feedback mechanisms. The phenomenological implications for dilepton production are discussed in Sec.~\eqref{3}. Finally, Sec.~\eqref{4} is devoted to the concluding remarks.

\medskip
\noindent\textbf{Notations and Conventions:} We work in natural units with $\hbar = k_B = c = \mu_0 = \epsilon_0 = 1$, and adopt the metric signature $g^{\mu\nu} = \mathrm{diag}(-,+,+,+)$ throughout this work. We define symmetrization and antisymmetrization of tensors as $
A^{(\mu}B^{\nu)} \equiv \frac{1}{2}\left(A^\mu B^\nu + A^\nu B^\mu\right), \quad
A^{[\mu}B^{\nu]} \equiv \frac{1}{2}\left(A^\mu B^\nu - A^\nu B^\mu\right).$



\section{Relativistic BDNK Magnetohydrodynamics}{\label{1}}
\subsection{Conservation Laws and Decomposition of Currents}
For a relativistic system, the macroscopic dynamics are formulated in terms of conserved currents, which encode the evolution of energy, momentum, electromagnetic fields, and conserved charges. In particular, we begin by introducing the conservation laws governing the energy-momentum tensor, the dual electromagnetic field strength tensor, and the number current.
The dynamics of the system are therefore determined by the conservation equations
\begin{align}
\nabla_\mu T^{\mu\nu} = 0 \,, \quad
\nabla_\mu J^{\mu\nu} = 0 \,, \quad
\nabla_\mu N^\mu = 0 \,,
\label{eq:conservation_all}
\end{align}
where $T^{\mu\nu}$ is the symmetric energy-momentum tensor, $N^\mu$ is the conserved number current, and $J^{\mu\nu} = \epsilon^{\mu\nu\alpha\beta} F_{\alpha\beta}$ denotes the dual electromagnetic field strength tensor constructed from the electromagnetic field tensor $F_{\alpha\beta}$. The antisymmetry of $J^{\mu\nu}$ follows directly from its definition.
In a hydrodynamic description, these conserved quantities are expressed as an expansion in space-time gradients of the hydrodynamic fields~\cite{Kovtun:2019hdm},
\begin{align}
T^{\mu\nu} &= T^{\mu\nu}_{(0)} + T^{\mu\nu}_{(1)} + \cdots \,, \quad
J^{\mu\nu} = J^{\mu\nu}_{(0)} + J^{\mu\nu}_{(1)} + \cdots \,, \nonumber\\
N^{\mu} &= N^{\mu}_{(0)} + N^{\mu}_{(1)} + \cdots \,,
\end{align}
where the subscript $(0)$ denotes the ideal (perfect-fluid) contribution, while $(1)$ represents first-order dissipative and gradient corrections. Higher-order terms in the derivative expansion are systematically suppressed in the hydrodynamic regime.
Using the fluid four-velocity $u^\mu$, a unit timelike vector satisfying $u^\mu u_\mu = -1$, any tensorial structure can be decomposed into components parallel and orthogonal to the local rest frame defined by $u^\mu$. In particular, a symmetric rank-2 tensor, an antisymmetric tensor, and a vector can be decomposed as
\begin{align}
T^{\mu\nu} &= \mathcal{E}\, u^\mu u^\nu + \mathcal{P}\, \Delta^{\mu\nu}
+ \mathcal{Q}^\mu u^\nu + \mathcal{Q}^\nu u^\mu
+ \mathcal{T}^{\mu\nu} \,, \label{eq:tmn-gen}\\
J^{\mu\nu} &= u^\mu \mathcal{B}^{\nu} - u^\nu \mathcal{B}^{\mu}
+ \mathcal{D}^{\mu\nu} \,, \label{eq:Jmn-gen}\\
N^{\mu} &= \mathcal{N}\, u^\mu + \mathcal{N}^\mu \,. \label{eq:Nm-gen}
\end{align}

Here, $\Delta^{\mu\nu} = g^{\mu\nu} + u^\mu u^\nu$ is the projector orthogonal to $u^\mu$. The quantities $\mathcal{Q}^\mu$, $\mathcal{B}^\mu$, $\mathcal{N}^\mu$, $\mathcal{T}^{\mu\nu}$, and $\mathcal{D}^{\mu\nu}$ are all transverse to the fluid velocity, i.e., orthogonal to $u^\mu$. In addition, $\mathcal{T}^{\mu\nu}$ is symmetric and traceless, while $\mathcal{D}^{\mu\nu}$ is antisymmetric. The scalar and tensor components appearing above are defined through appropriate projections as
\begin{align}
\mathcal{E} &\equiv T^{\mu\nu} u_\mu u_\nu \,, \quad
\mathcal{P} \equiv \frac{1}{3} \Delta_{\mu\nu} T^{\mu\nu} \,, \quad
\mathcal{N} \equiv -u_\mu N^\mu \,, \nonumber\\
\mathcal{Q}^\mu &\equiv -\Delta^{\mu}_{\ \alpha} u_\beta T^{\alpha\beta} \,, \quad
\mathcal{B}^\mu \equiv J^{\mu\nu} u_\nu \,, \quad
\mathcal{N}^\mu \equiv \Delta^\mu_{\ \nu} N^\nu \,, \nonumber\\
\mathcal{T}^{\mu\nu} &\equiv \Delta^{\mu\nu}_{\alpha\beta} T^{\alpha\beta} \,, \quad
\mathcal{D}^{\mu\nu} \equiv \Delta^{[\mu}_ \alpha\Delta^{\nu]}_\beta J^{\alpha\beta} \,.
\end{align}

Here, $\Delta^{\mu\nu}_{\alpha\beta}$ denotes the symmetric traceless projector orthogonal to $u^\mu$, defined as $\Delta^{\mu\nu}_{\alpha\beta}
= \frac{1}{2}
\left(
\Delta^\mu_{\ \alpha}\Delta^\nu_{\ \beta}
+ \Delta^\nu_{\ \alpha}\Delta^\mu_{\ \beta}
\right)
- \frac{1}{3}\Delta^{\mu\nu}\Delta_{\alpha\beta} \,.$
In the presence of a magnetic field, it is convenient to introduce an additional spacelike unit vector $h^\mu$ satisfying $h^\mu h_\mu = 1$ and $u_\mu h^\mu = 0$. This allows us to define a projector orthogonal to both $u^\mu$ and $h^\mu$ as $\Delta^{\mu\nu}_\perp = g^{\mu\nu} + u^\mu u^\nu - h^\mu h^\nu \,.$
With this additional structure, the conserved currents admit a refined decomposition,
\begin{align}
T^{\mu\nu} &= \mathcal{E} u^\mu u^\nu
+ \left(\mathcal{P} - \frac{1}{2}\mathcal{S}\right)\Delta^{\mu\nu}_\perp
+ \left(\mathcal{P} + \mathcal{S}\right) h^\mu h^\nu \nonumber\\
&\quad + \mathcal{Q}_\parallel h^{(\mu}u^{\nu)}
+ \mathcal{Q}^{(\mu}_\perp u^{\nu)}
+ \mathcal{T}^{(\mu}_\perp h^{\nu)}
+ \mathcal{T}^{\mu\nu}_\perp \,, \label{eq:Tmn-de-uh}\\
J^{\mu\nu} &= \mathcal{B}_\parallel u^{[\mu}h^{\nu]}
+ u^{[\mu}\mathcal{B}^{\nu]}_\perp
+ \mathcal{D}^{[\mu}_\perp h^{\nu]}
+ \mathcal{D}^{\mu\nu}_\perp \,, \label{eq:Jmn-de-uh}\\
N^\mu &= \mathcal{N} u^\mu + \mathcal{N}_\parallel h^\mu + \mathcal{N}^\mu_\perp \,. \label{eq:N-de-uh}
\end{align}

The additional coefficients appearing in this decomposition are obtained by projections along $u^\mu$, $h^\mu$, and the transverse subspace. For instance,
\begin{align}
\mathcal{S} &\equiv h_\mu h_\nu \mathcal{T}^{\mu\nu} \,, \quad
\mathcal{Q}_\parallel \equiv h_\mu \mathcal{Q}^\mu \,, \quad
\mathcal{Q}^\mu_\perp \equiv \Delta^{\mu\nu}_\perp \mathcal{Q}_\nu \,, \nonumber\\
\mathcal{T}^\mu_\perp &\equiv \Delta^{\mu\alpha}_\perp h^\beta \mathcal{T}_{\alpha\beta} \,, \quad
\mathcal{D}^\mu_\perp \equiv \mathcal{D}^{\mu\nu} h_\nu \,, \nonumber\\
\mathcal{D}^{\mu\nu}_\perp &\equiv \frac{1}{2}
\Delta_\perp^{\alpha[\mu}\Delta_\perp^{\nu]\beta}\mathcal{D}_{\alpha\beta} \,, \nonumber\\
\mathcal{T}^{\mu\nu}_\perp &\equiv
\lb \Delta_\perp^{\alpha(\mu}\Delta_\perp^{\nu)\beta}-\frac{1}{2}\Delta_\perp^{\mu\nu}\Delta_\perp^{\alpha\beta}\rb \mathcal{T}_{\alpha\beta} \,.
\end{align}

Similarly, $\mathcal{B}_\parallel$, $\mathcal{B}^\mu_\perp$, $\mathcal{N}_\parallel$, and $\mathcal{N}^\mu_\perp$ are obtained by projecting $\mathcal{B}^\mu$ and $\mathcal{N}^\mu$ along $h^\mu$ and $\Delta^{\mu\nu}_\perp$, respectively.
At zeroth order in gradients, the conserved currents take the equilibrium form~\cite{Hoult:2024qph}
\begin{align}
T^{\mu\nu}_{(0)} &= \left(T \frac{\partial p}{\partial T}
+ \mu \frac{\partial p}{\partial \mu}
+ \mu_\phi \frac{\partial p}{\partial \mu_\phi}\right) u^\mu u^\nu
+ p g^{\mu\nu}
- \mu_\phi \rho_\phi h^\mu h^\nu \,,\\
J^{\mu\nu}_{(0)} &= \rho_\phi \left(u^\mu h^\nu - u^\nu h^\mu\right) \,,\\
N^\mu_{(0)} &= \left(\frac{\partial p}{\partial \mu}\right) u^\mu \,.
\end{align}

Using the hydrodynamic variables $\{T, u^\mu, \mu, \mu_\phi, h^\mu\}$, one can systematically construct all allowed first-order gradient structures. These include parity-even scalars $(s_i)$, parity-odd scalars $(p_i)$, transverse vectors $(Y^\mu_i)$ and $(\Sigma^\mu_i)$, as well as symmetric and antisymmetric transverse tensors. Their explicit forms are
\begin{align}
    s_1 &= \frac{\dot{T}}{T} \,, \quad
s_2 = \nabla_\mu u^\mu \,, \quad
s_3 = u^\mu \partial_\mu \left(\frac{\mu}{T}\right) \,, \nn
s_4 &= h^\mu h^\nu \nabla_\mu u_\nu \,, \quad
s_5 = \frac{T}{\mu_\phi} u^\mu \partial_\mu \left(\frac{\mu_\phi}{T}\right) \,, \nn p_1 &= h_\mu \left(\frac{\Delta^{\mu\nu}\partial_\nu T}{T} + \dot{u}^\mu\right) \,, \quad
p_2 = \frac{1}{T\rho_\phi} \nabla_\mu (T\rho_\phi h^\mu) \,, \nn
p_3 &= h^\mu \partial_\mu \left(\frac{\mu}{T}\right) \,,\quad Y^\mu_1 = T \Delta^{\mu\nu}_\perp \left(\partial_\nu \frac{\mu_\phi}{T}
- \frac{\mu_\phi}{T} h^\alpha \nabla_\alpha h_\nu \right) \,,\nn
Y^\mu_2 &= T \Delta^{\mu\nu}_\perp \left(\frac{\Delta^\alpha_{\ \nu}\partial_\alpha T}{T} + \dot{u}_\nu\right) \,, \quad
Y^\mu_3 = T \Delta^{\mu\nu}_\perp \partial_\nu \left(\frac{\mu}{T}\right) \,, \nn
\Sigma^\mu_1 &= 2\Delta^{\mu\alpha}_\perp h^\nu \nabla_{(\alpha} u_{\nu)}  \,, \quad
\Sigma^\mu_2 = 2\Delta^{\mu\alpha}_\perp u^\nu \nabla_{[\alpha} h_{\nu]} \,,\nn
\sigma^{\mu\nu}_\perp &= \left(
2\Delta_\perp^{\alpha(\mu}\Delta_\perp^{\nu)\beta}
- \Delta_\perp^{\mu\nu}\Delta_\perp^{\alpha\beta}
\right)\nabla_\alpha u_\beta \,, \nn
Z^{\mu\nu} &= 2\mu_\phi \Delta_\perp^{\mu\rho}\Delta_\perp^{\nu\sigma}
\nabla_{[\rho} h_{\sigma]} \,.
\end{align}

Finally, the most general first-order constitutive relations for the conserved currents can be written as
\begin{align}\label{eq:full-expan-coe}
    \mathcal{E} &= \epsilon + \sum_{i=1}^5 \varepsilon_i s_i \,, \quad
\mathcal{P} = p - \frac{1}{3}\mu_\phi \rho_\phi + \sum_{i=1}^5 \pi_i s_i \,, \nn
\mathcal{S} &= -\frac{2}{3}\mu_\phi \rho_\phi + \sum_{i=1}^5 \sigma_i s_i \,, \quad
\mathcal{B}_\parallel = \rho_\phi + \sum_{i=1}^5 \beta_{\parallel i} s_i \,,\nn
\mathcal{N} &= n + \sum_{i=1}^5 \nu_i s_i \,, \quad
\mathcal{Q}_\parallel = \sum_{i=1}^3 \theta_{\parallel i} p_i \,, \quad
\mathcal{N}_\parallel = \sum_{i=1}^3 \gamma_{\parallel i} p_i \,, \nn
\mathcal{Q}^\mu_\perp &= \sum_{i=1}^3 \theta_{\perp i} Y^\mu_i \,, \quad
\mathcal{D}^\mu_\perp = \sum_{i=1}^3 \rho_{\perp i} Y^\mu_i \,,\quad \mathcal{N}^\mu_\perp = \sum_{i=1}^3 \gamma_{\perp i} Y^\mu_i \,, \nn
\mathcal{T}^\mu_\perp &= \sum_{i=1}^2 \tau_i \Sigma^\mu_i \,, \quad
\mathcal{B}^\mu_\perp = \sum_{i=1}^2 \beta_{\perp i} \Sigma^\mu_i \,,\quad \mathcal{T}^{\mu\nu}_\perp = -\eta_\perp \sigma^{\mu\nu}_\perp \,, \nn
\mathcal{D}^{\mu\nu}_\perp &= -r_\parallel Z^{\mu\nu} \,.
\end{align}
Here, $\epsilon$, $p$, $\rho_\phi$, and $n$ denote the equilibrium thermodynamic quantities, while the coefficients $\varepsilon_i$, $\pi_i$, $\sigma_i$, $\beta_{\parallel i}$, $\nu_i$, $\theta_{\parallel i}$, $\gamma_{\parallel i}$, $\theta_{\perp i}$, $\rho_{\perp i}$, $\gamma_{\perp i}$, $\tau_i$, $\beta_{\perp i}$, $\eta_\perp$, and $r_\parallel$ represent first-order transport parameters. In total, the theory contains 46 independent transport coefficients, although not all correspond to independent physical transport processes~\cite{Hoult:2024qph}.

\subsection{BDNK Magnetohydrodynamic Equations in a Boost-Invariant System
}
For the numerical implementation of solving the set of conservation equations, we employ the Bjorken flow approximation~\cite{Bjorken:1982qr}, which assumes boost-invariant longitudinal expansion and provides a reliable description of the early-time dynamics near mid-rapidity in relativistic heavy-ion collisions. In this framework, all macroscopic quantities depend solely on the proper time $\tau$, effectively reducing the system to a $(0+1)$D evolution.
We work in Milne coordinates, where $\tau=\sqrt{t^2-z^2}$ denotes the proper time and $\eta=\frac{1}{2}\ln\!\left(\frac{t+z}{t-z}\right)$ is the space-time rapidity. Under boost invariance and transverse homogeneity, the fluid velocity takes the comoving form $u^\mu=(1,0,0,0)$, and spatial gradients in the transverse and rapidity directions are neglected. For simplicity, the magnetic field is chosen along the $y$-direction, $h^\mu=(0,0,1,0)$.
With these assumptions, the conservation laws reduce to a coupled system of nonlinear ordinary differential equations in $\tau$, which can be solved numerically in a controlled manner~\cite{Gangadharan:2023yzb}. This setup provides a transparent baseline to investigate the interplay between hydrodynamic expansion and electromagnetic fields. In the ideal limit, it reproduces the standard Bjorken scaling solution, serving as a reference to quantify deviations induced by transport coefficients and backreaction effects.
Under these symmetry constraints, the number of independent transport coefficients reduces to a manageable subset ({16} non-zero and in the present setup), and the conservation equations simplify to
\begin{align} {\label{eq:all}}
\partial_\tau \mathcal{E} 
+ \frac{1}{\tau}\left( \mathcal{E} + \mathcal{P} - \frac{1}{2}\mathcal{S} \right)
- \frac{1}{\tau^2}\eta_\perp &= 0 \,, \\
\partial_y \left( \mathcal{P} + \mathcal{S} \right) &= 0 \,, \\
\partial_x \left[
\left( \mathcal{P} - \frac{1}{2}\mathcal{S} \right)
+ \frac{1}{\tau}\eta_\perp
\right] &= 0 \,, \\
\partial_\eta \left[
\frac{1}{\tau^2}\left( \mathcal{P} - \frac{1}{2}\mathcal{S} \right)
- \frac{1}{\tau^3}\eta_\perp
\right] &= 0 \,, \\
\partial_\tau \mathcal{B}_\parallel + \frac{1}{\tau}\mathcal{B}_\parallel &= 0 \,, \\
\partial_y \mathcal{B}_\parallel &= 0 \,, \\
\Delta^{k}_{\perp y}\left(
\partial_\tau \mathcal{B}_\parallel + \frac{1}{\tau}\mathcal{B}_\parallel
\right) &= 0 \,, \\
\partial_\tau \mathcal{N} + \frac{1}{\tau}\mathcal{N} &= 0 \,.{\label{eq:allend}}
\end{align}
All quantities appearing above are defined in Eq.~\eqref{eq:full-expan-coe}, include both ideal and first-order gradient contributions. The thermodynamic relations are given by $p = p_{\rm m} - \frac{B^2}{2} + \frac{B^2}{\mu_B}, \quad
\epsilon = \epsilon_{\rm m} + \frac{1}{2}B^2, \quad
h^\mu = \frac{B^\mu}{B}, \quad
\rho_\phi = B, \quad
\mu_\phi = \frac{B}{\mu_B}.
$ Here, $\epsilon_{\rm m}$ and $p_{\rm m}$ denote the matter (fluid) energy density and pressure, respectively, i.e. the contributions arising purely from the strongly interacting medium in the absence of electromagnetic fields. The quantity $B^\mu$ represents the magnetic field four-vector with magnitude $B=\sqrt{B^\mu B_\mu}$. The parameter $\mu_B$ denotes the magnetic permeability, $\rho_\phi$ is identified with the magnetic field strength in the comoving frame, and $\mu_\phi$ is the associated magnetic chemical potential-like variable.
We now collect the complete set of equations obtained from the conservation laws, summarized in Eqs.~\eqref{eq:all}--\eqref{eq:allend}. This yields a system of nine coupled equations derived from the fundamental conservation relations introduced earlier in Eq.~\eqref{eq:conservation_all}. A closer inspection of this system shows that the essential dynamics is governed by the temporal evolution of the energy density, number density, and magnetic field. The remaining equations do not introduce additional nontrivial spatial structure, reflecting the boost-invariant and effectively spatially homogeneous nature of the present setup.
In what follows, we therefore focus on the reduced set of evolution equations governing the time dynamics. For consistency and clarity, we express all transport coefficients in terms of their temperature scaling as dictated by dimensional analysis. We parametrize
\begin{align}
\varepsilon_i &= \varepsilon_{0i}\, T^3 \,, \qquad
\pi_i = \pi_{0i}\, T^3 \,, \qquad
\sigma_i = \sigma_{0i}\, T^3 \,, \nonumber\\[4pt]
\beta_{\parallel i} &= \beta_{\parallel 0i}\, T \,, \qquad
\nu_i = \nu_{0i}\, T^2 \,,
\end{align}
{here $\varepsilon_{0i},\pi_{0i},\sigma_{0i},\beta_{\para 0i},\nu_{0i}$ are all constants.}
We further adopt thermodynamic inputs for the medium with an ideal equation of state~\cite{Armas:2022wvb},
\begin{align}
\begin{aligned}
\epsilon_{\rm m} &= 74.1 \frac{N_c^2}{2\pi^2} T^4 \,, \qquad
p_{\rm m} = \frac{1}{3}\epsilon_{\rm m} \,, \\
s &= 99.4 \frac{N_c^2}{2\pi^2} T^3 \,, \qquad
\eta_\perp = \frac{s}{4\pi} \,.
\end{aligned}
\end{align}
where $s$ denotes the entropy density and $N_c$ is the number of colours, which we set to $N_c=1$ for simplicity. Throughout this work, we also take $\mu_B = 1$, corresponding to neglecting magnetic polarization effects of the medium. The reduced set of equations is therefore given by
\begin{equation}
\begin{aligned}
4aT^3\dot T + B\dot B
+ \frac{1}{\tau}
\left(
\frac{4}{3}aT^4 + B^2
\right)- \frac{b}{4\pi}\frac{T^3}{\tau^2}& \\[4pt]
+ \varepsilon_{02}
\left(
\frac{3T^2\dot T}{\tau}
- \frac{T^3}{\tau^2}
\right)
+ 
(\varepsilon_{01}+\pi_{01}-\tfrac{1}{2}\sigma_{01})\, \frac{T^2 \dot T}{\tau}
 &\\[4pt]
+ 
(\varepsilon_{05}+\pi_{05}-\tfrac{1}{2}\sigma_{05})\, 
\left(
\frac{T^3\dot B}{\tau B} - \frac{T^2\dot T}{\tau}
\right)&\\[4pt]
+ 
(\varepsilon_{02}+\pi_{02}-\tfrac{1}{2}\sigma_{02})\, \frac{T^3}{\tau^2}
= 0 \,&.\\
\dot B
+ \frac{B}{\tau}
+ \lb \beta_{\parallel 01}+\beta_{\parallel 02}-\beta_{\parallel 05} \rb \, \frac{\dot T}{\tau}
+ \frac{1}{\tau}\beta_{\parallel 05}\, \frac{T \dot B}{\tau B}
= 0 \,, &\\[6pt]
\dot n
+ \frac{n}{\tau} 
+ \lb \nu_{01} + 2\nu_{02} - \nu_{05}\rb \,\frac{ T\dot T}{\tau}
+ \nu_{05}\,  \frac{T^2\dot B}{\tau B}
= 0 \,, &
\end{aligned}
\label{eq:coupled_evolution}
\end{equation}
where $a= \frac{74.1}{2 \pi^2}$ and $b =  \frac{99.4 }{2\pi^2}$. In solving the coupled equations, we restrict ourselves to first-order dynamics in proper time. Accordingly, we retain only terms linear in the time derivatives and neglect all higher-order contributions, such as $\ddot{T}$, $\ddot{B}$, mixed terms like $\dot{T}\dot{B}$, and quadratic derivative terms $\dot{T}^2$ and $\dot{B}^2$. This approximation is consistent with a first-order gradient expansion and ensures a controlled and tractable evolution framework. Also, throughout this work we neglect shear effects by setting \( b = 0 \) in the temperature evolution, and correspondingly ignore their backreaction on the magnetic field dynamics. This amounts to omitting shear viscosity contributions within the present setup. While their inclusion would modify the quantitative results, we do not expect it to qualitatively alter the observed trends, such as relative enhancements or suppressions, which remain robust.
\section{Systematic Study of the effect of expansion scalars}{\label{2}}
The set of coupled evolution equations solved in this work is presented in Eq.~\eqref{eq:coupled_evolution}. 
The dynamics are governed by a collection of transport parameters, namely 
$\nu_i$, $\beta_{\parallel i}$, $\varepsilon_{i}$, $\pi_{i}$, $\sigma_{i},$ and $\eta_\perp$, 
which encode the various dissipative and electromagnetic contributions within the first-order gradient expansion. 
Here, the index $i$ labels the independent scalar structures entering the gradient expansion. 
It runs over the allowed structures (effectively $i=1\text{--}5$), with $s_3=0$ due to the assumption of {vanishing chemical potential $(\mu=0)$} and $s_4 = 0$ as a consequence of the chosen field configuration and flow profile. 
The numerical values of these parameters are taken to lie within the range $(-1,1)$. To disentangle the coupled dynamics of the temperature $T(\tau)$, magnetic field $B(\tau)$, and number density $n(\tau)$, we systematically vary the transport parameters in a controlled and physically motivated manner. 
By selectively activating different subsets, we isolate distinct physical mechanisms such as the influence of the magnetic field on the temperature and its backreaction on the magnetic field evolution, and assess their combined impact on the number density in a transparent way.

\subsection{Baseline: No Feedback (Ideal Limit)}
\begin{figure}[h]
    \centering
    \includegraphics[width=\linewidth]{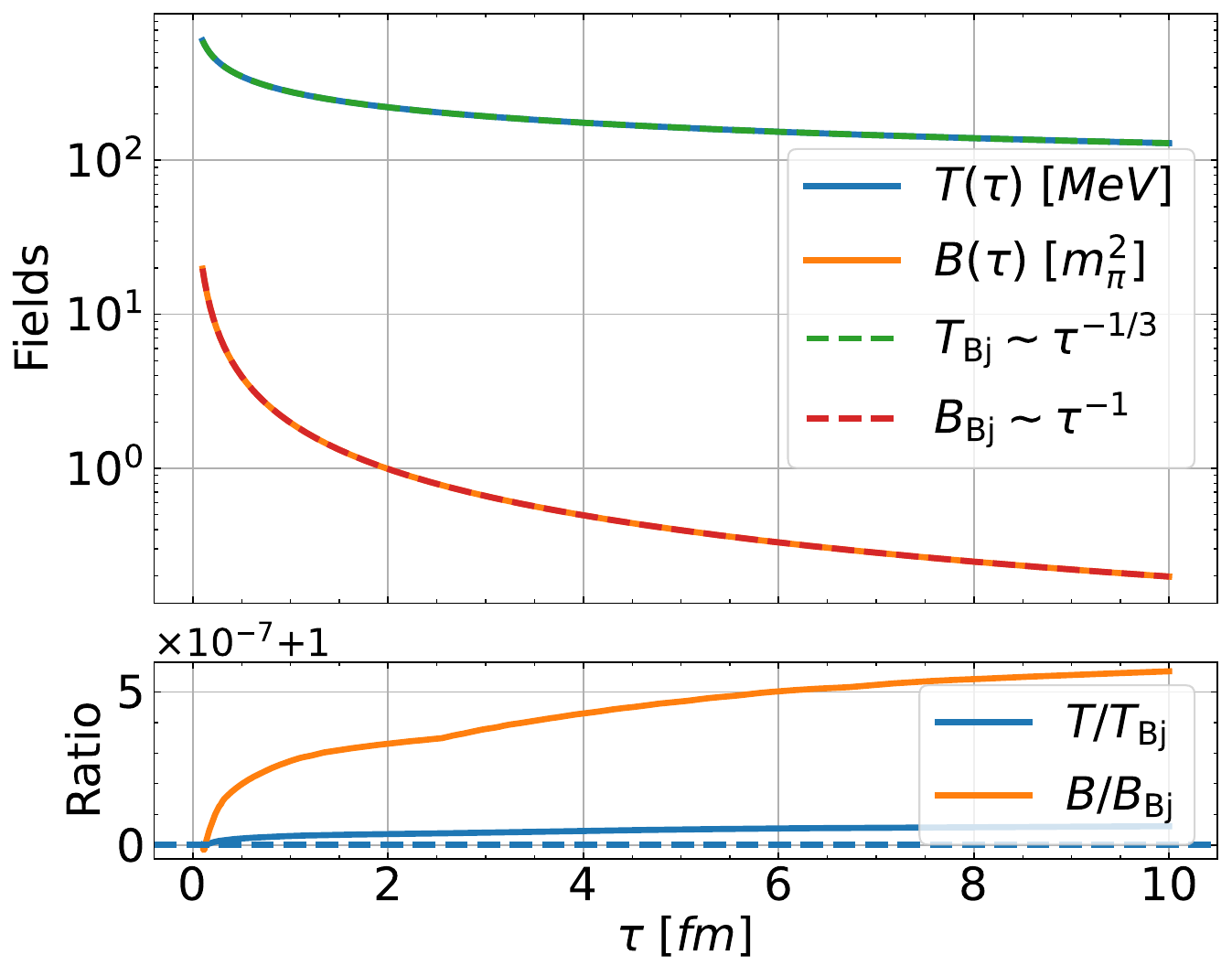}
    \caption{Evolution of the temperature and magnetic field with all expansion coefficients set to zero, for initial conditions $T_{\text{in}} = 600~\text{MeV}$ and $B_{\text{in}} = 20~m_{\pi}^2$.}
    \label{fig:solution_base}
\end{figure}

We begin with the baseline scenario in which all first-order derivative expansion terms are set to zero. In this limit, the system reduces to the well-known ideal Bjorken solution~\cite{Roy:2015kma}. This expectation is confirmed in Fig.~\eqref{fig:solution_base}, where the evolution of the temperature (blue) and magnetic field (orange) closely follows the corresponding ideal Bjorken scaling laws $T(\tau)\sim \tau^{-1/3}\,, \ B(\tau)\sim \tau^{-1}$. 
The ratios shown in the lower panel remain extremely close to unity, with deviations at the level of $\mathcal{O}(10^{-7})$. 
For this setup, we initialize the system with a temperature $T_{\mathrm{in}} = 600~\mathrm{MeV}$ and a magnetic field $B_{\mathrm{in}} = 20~m_{\pi}^2$. In the absence of expansion parameters, the magnetic field remains entirely decoupled from the thermal evolution. 
While the temperature does not directly influence the magnetic field in this limit, the latter still enters the energy conservation equation and thereby affects the evolution of $T(\tau)$. 
The influence of the magnetic field on the temperature evolution arises from two competing contributions: the term proportional to $B\,\dot{B}$ and the one proportional to $B^2/\tau$. The former reflects the temporal variation of the magnetic field and, although $B\,\dot{B}$ is negative in an expanding system where $B \sim \tau^{-1}$, it effectively reduces the magnitude of $\dot{T}$, thereby moderating the cooling of the system. In contrast, the $B^2/\tau$ term originates from the magnetic pressure and reinforces the dilution caused by the expansion, leading to a more rapid decrease in temperature. The evolution of $T(\tau)$ is thus governed by a competition between these opposing effects. For the characteristic scaling $B \sim \tau^{-1}$, as expected within the Bjorken framework, both contributions scale identically but with opposite signs, resulting in an approximate cancellation. As a consequence, the net effect of the magnetic field becomes subleading, and the temperature evolution closely follows the ideal Bjorken scaling.

\subsection{Effect of Temperature on Magnetic Field}
 First, the coupling between temperature gradients and the magnetic field evolution is introduced through the transport parameters $\beta_{\parallel i}$, which encode how thermal gradients enter the magnetic field evolution equation and thereby mediate the influence of the temperature sector on the electromagnetic dynamics. 
In this setup, we activate only these transport parameters, $\beta_{\parallel i} \neq 0$, while setting all other transport parameters to zero, i.e., $\varepsilon_{0i} = \pi_{0i} = \sigma_{0i} = 0 \,.$
\begin{figure}[h]
    \centering
    \includegraphics[width=\linewidth]{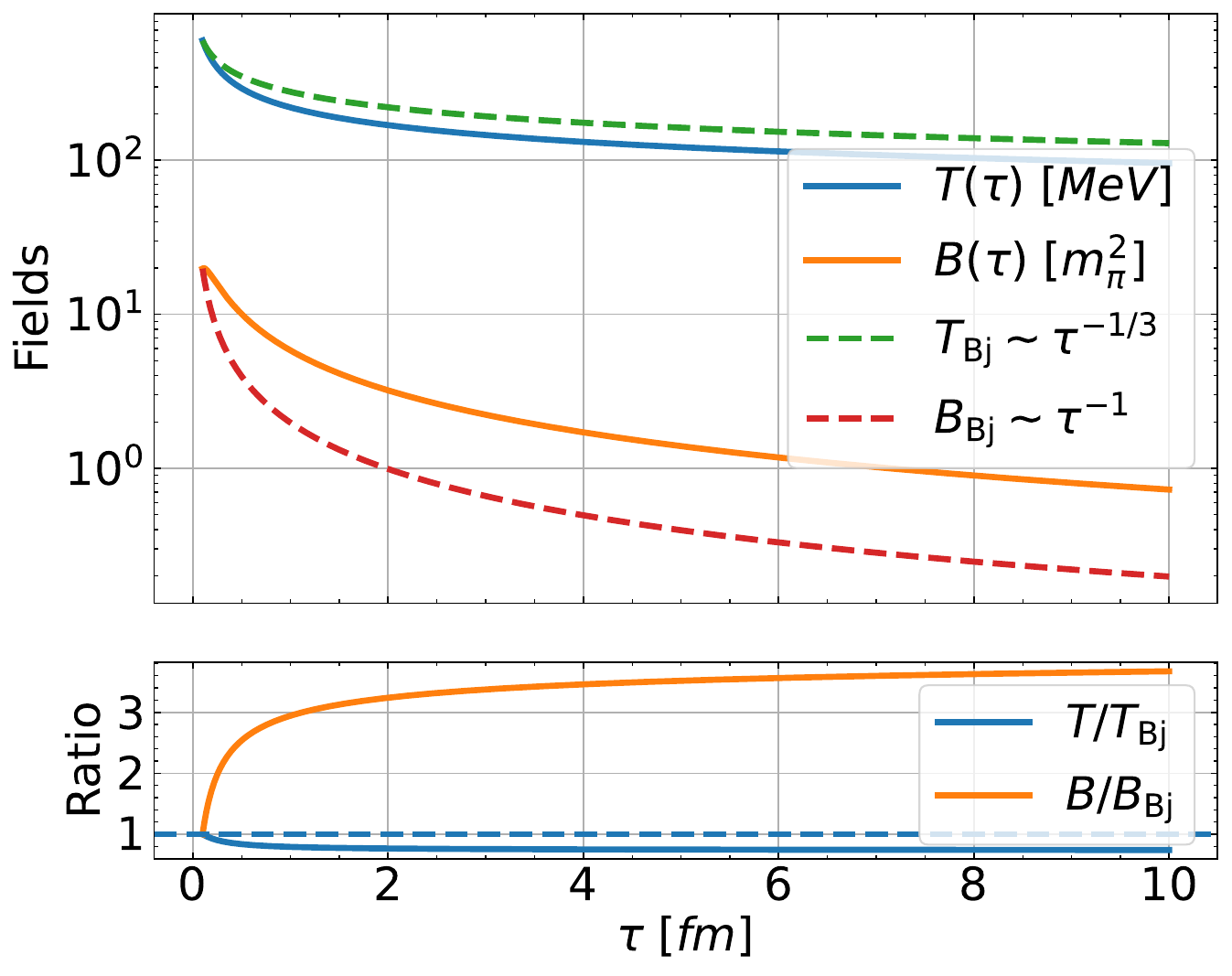}
    \includegraphics[width=\linewidth]{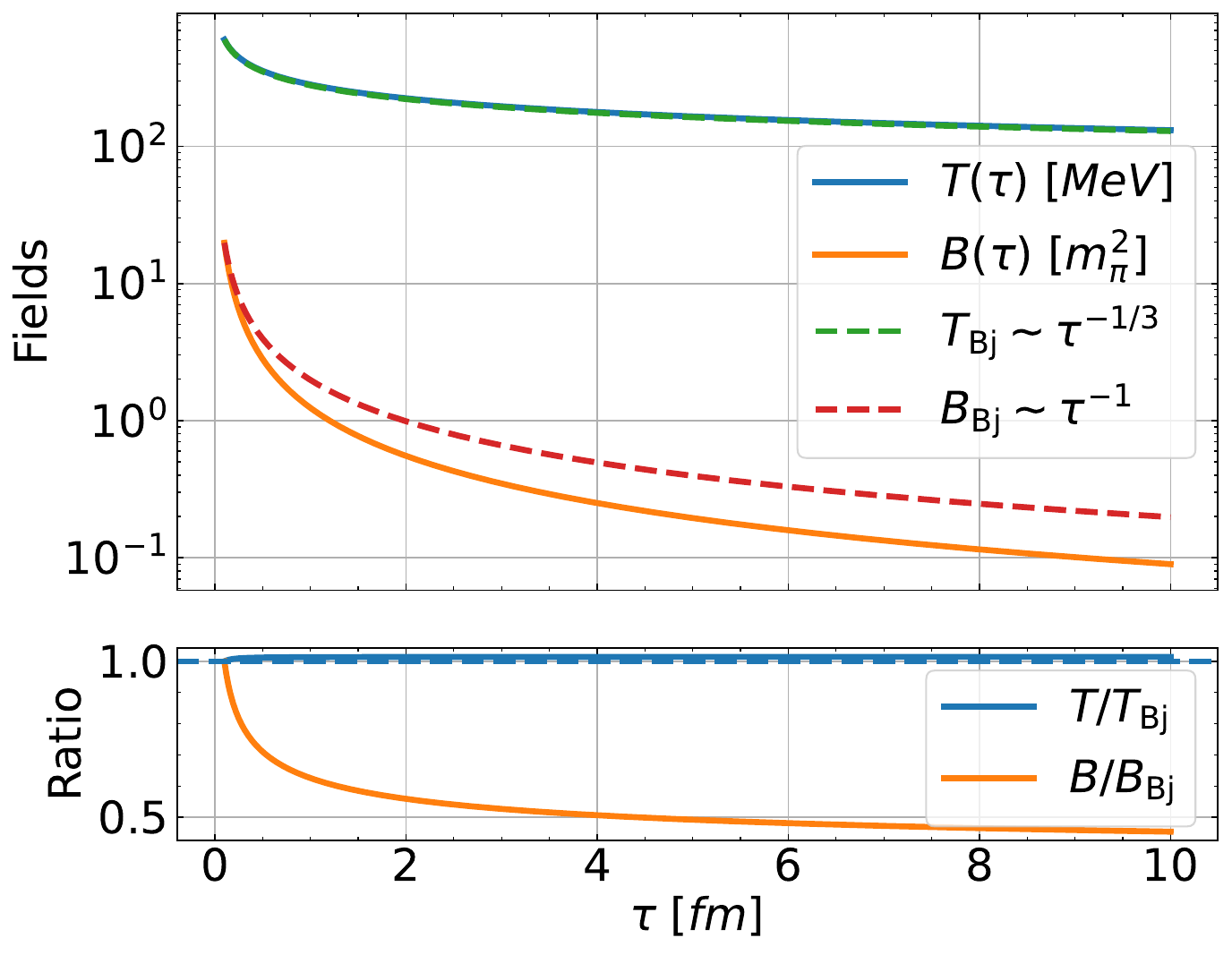}
    \caption{Evolution of the temperature and magnetic field with all expansion coefficients set to zero except $\beta_{\parallel 0i}$. The system is initialized with $T_{\text{in}} = 600~\mathrm{MeV}$ and $B_{\text{in}} = 20~m_{\pi}^2$. The upper panel corresponds to $\left(\beta_{\parallel 01}, \beta_{\parallel 02}, \beta_{\parallel 05}\right) = (0.6,\,0.4,\,0.1)$, while the lower panel corresponds to $\left(\beta_{\parallel 01}, \beta_{\parallel 02}, \beta_{\parallel 05}\right) = (-0.1,\,-0.05,\,-0.03)$.}
    \label{fig:solution_temp_on_mag}
\end{figure}
This choice isolates the effect of temperature on the magnetic field dynamics. 
The resulting evolution is shown in Fig.~\eqref{fig:solution_temp_on_mag}. The upper panel corresponds to positive values of the $\beta_{\parallel i}$ parameters. In this case, the temperature-driven coupling acts to slow down the decay of the magnetic field ( orange curve ) compared to the ideal Bjorken scaling. This modification in the magnetic field evolution, in turn, feeds back into the temperature dynamics through the competing $B\,\dot{B}$ and $B^2$ contributions. We observe that the enhanced persistence of the magnetic field leads to a slightly faster cooling of the system, indicating that the dissipative contribution from the $B\,\dot{B}$ term becomes subdominant over the magnetic pressure term proportional to $B^2$.
{In} comparison with the {ideal} Bjorken {case, it} shows that the modification in the magnetic field evolution is significantly more pronounced—of the order of a few times larger—than the corresponding change in the temperature profile, which remains comparatively moderate.
The lower panel illustrates the same setup but with negative values of $\beta_{\parallel}$. In this case, the effect is qualitatively reversed: the magnetic field decays faster, and the corresponding feedback on the temperature evolution follows the opposite trend. The specific magnitudes of the negative coefficients differ from the positive case, as identical values lead to numerical instabilities in the present framework; this choice is therefore guided purely by the requirement of obtaining stable and consistent solutions.
 
\subsection{Effect of Magnetic Field on Temperature}
\begin{figure}
    \centering
    \includegraphics[width=\linewidth]{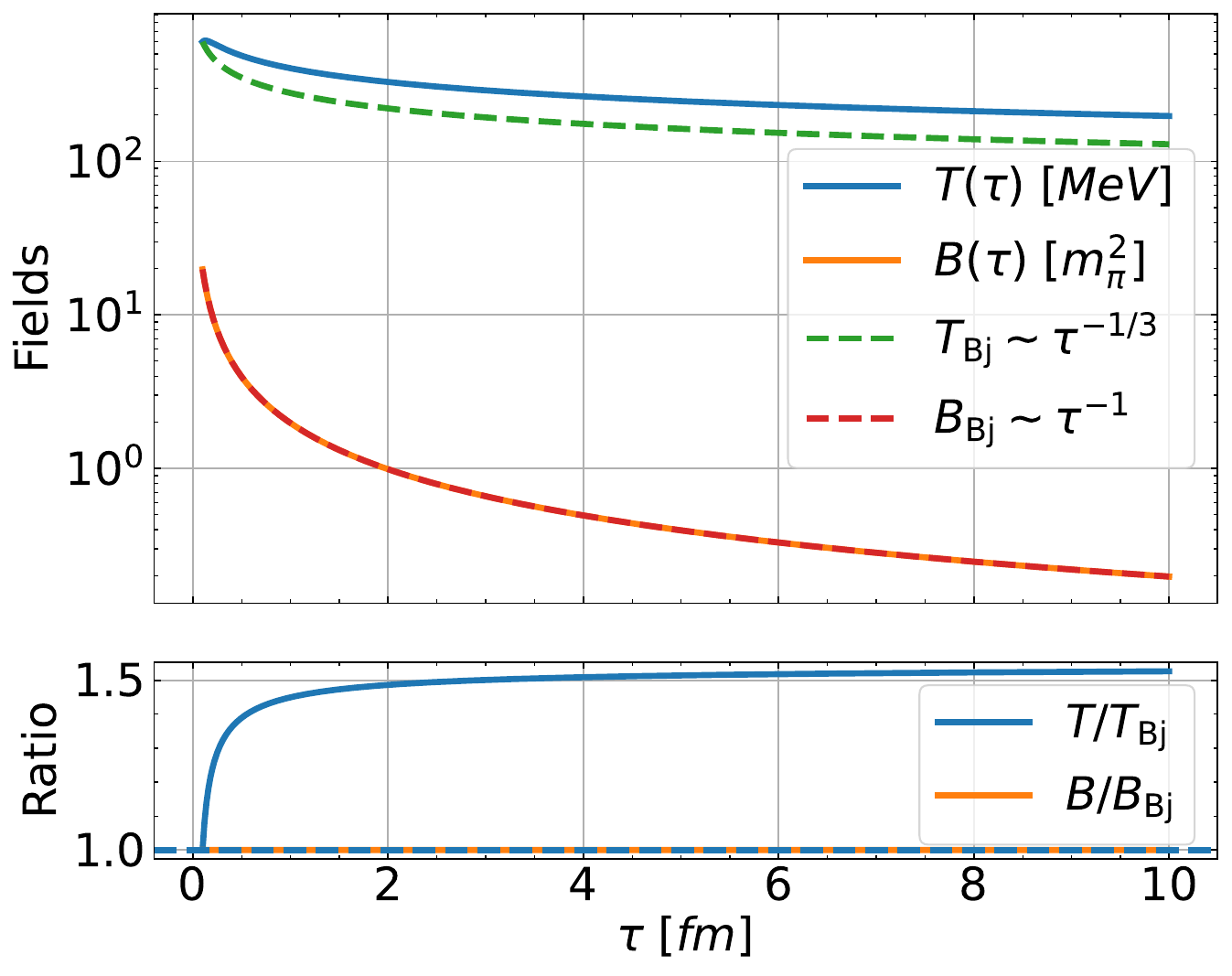}
    \includegraphics[width=\linewidth]{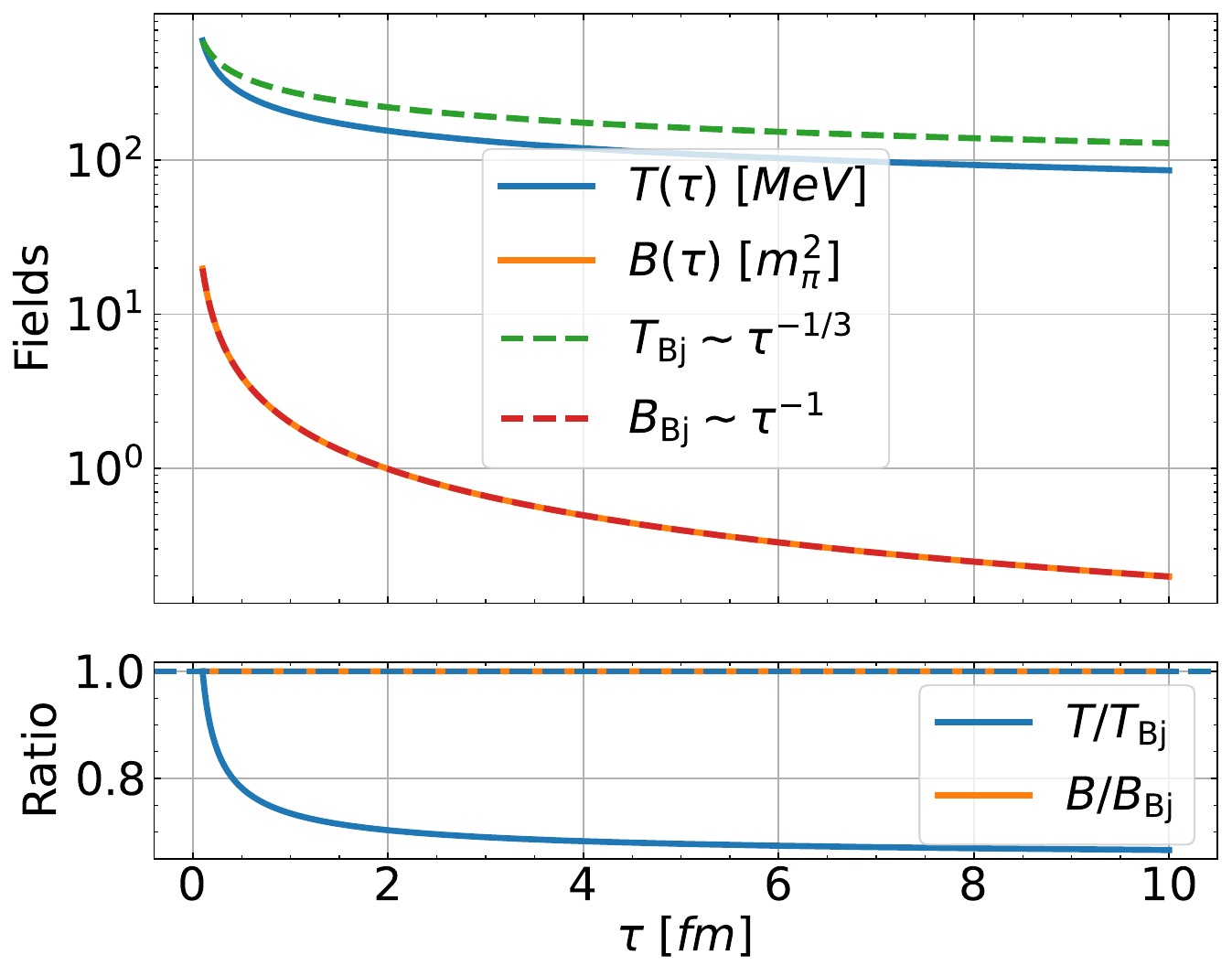}
    \caption{Evolution of the temperature and magnetic field with only the expansion coefficients $\varepsilon_{05}$, $\pi_{05}$, and $\sigma_{05}$ taken to be non-zero. The system is initialized with $T_{\text{in}} = 600~\mathrm{MeV}$ and $B_{\text{in}} = 20\,m_{\pi}^2$. The upper panel corresponds to $(\varepsilon_{05}, \pi_{05}, \sigma_{05}) = (0.99,\,0.99,\,0.03)$, while the lower panel corresponds to $(\varepsilon_{05}, \pi_{05}, \sigma_{05}) = (-0.99,\,-0.99,\,-0.03)$.}
    \label{fig:solution_mag_on_temp}
\end{figure}
{Now to see the feedback of the magnetic field on temperature, we activate the transport parameters that couple the magnetic field and temperature, i.e., $\varepsilon_{05}, \, \pi_{05}, \, \sigma_{05} \neq 0 \, $, and all other transport parameters are set to zero.}
{By this} setup {one can } isolate the pure feedback of the magnetic field on the thermal sector, while removing any influence of temperature gradients on the magnetic field evolution itself. Consequently, the temperature dynamics acquire additional magnetic contributions beyond the baseline coupling discussed earlier. Although a coupling is present even when the scalar coefficients vanish, its effect in that case is comparatively subtle.
The corresponding evolution is shown in Fig.~\eqref{fig:solution_mag_on_temp}. In the upper panel, 
{we choose positive transport parameters}, here the temperature exhibits a slower decay compared to the ideal Bjorken scaling. This behavior reflects a competition between temperature and magnetic field gradient contributions entering via the coupling. 
In this regime, the temperature evolution is primarily influenced by the $B\dot{B}$ contribution, relative to the competing $B^2$ term and the additional $T\dot{T}$ structure on the left-hand side of the evolution equation. Since $B\dot{B}<0$ for an expanding system, this term acts as a feedback from the decaying magnetic energy into the thermal sector. As a result, it reduces the effective cooling rate, leading to a slower decay of the temperature compared to the case where the magnetic coupling is absent or subdominant.
The magnetic field evolution on the other hand remains essentially unchanged and continues to follow the Bjorken scaling, reflecting the absence of any feedback from the thermal sector to the magnetic field. In the ratio plot (in the upper panel), we see the temperature is enhanced by approximately a factor of $\sim 1.5$ relative to the ideal case.
When the signs of the non-zero coefficients are reversed, as shown in the lower panel, the behavior is correspondingly inverted, leading to a faster-than-Bjorken decay of the temperature.
Another important observation is that, although the strength of this magnetic-to-temperature coupling is comparable to that of the previously studied temperature-to-magnetic-field coupling, its quantitative impact is noticeably weaker. In particular, the modification induced in the temperature evolution due to magnetic feedback is smaller than the corresponding change in the magnetic field driven by temperature gradients. This asymmetry is evident from a comparison of the ratio plots in Fig.~\eqref{fig:solution_temp_on_mag} and Fig.~\eqref{fig:solution_mag_on_temp}.

\subsection{Fully Coupled MHD System}
{Now to see the evolution of a fully coupled MHD system, we activated all the transport parameters, i.e., $\varepsilon_{0i}, \, \pi_{0i}, \, \sigma_{0i}, \, \beta_{\parallel 0i} \neq 0,
.$}
\begin{figure}[h]
    \centering
    \includegraphics[width=\linewidth]{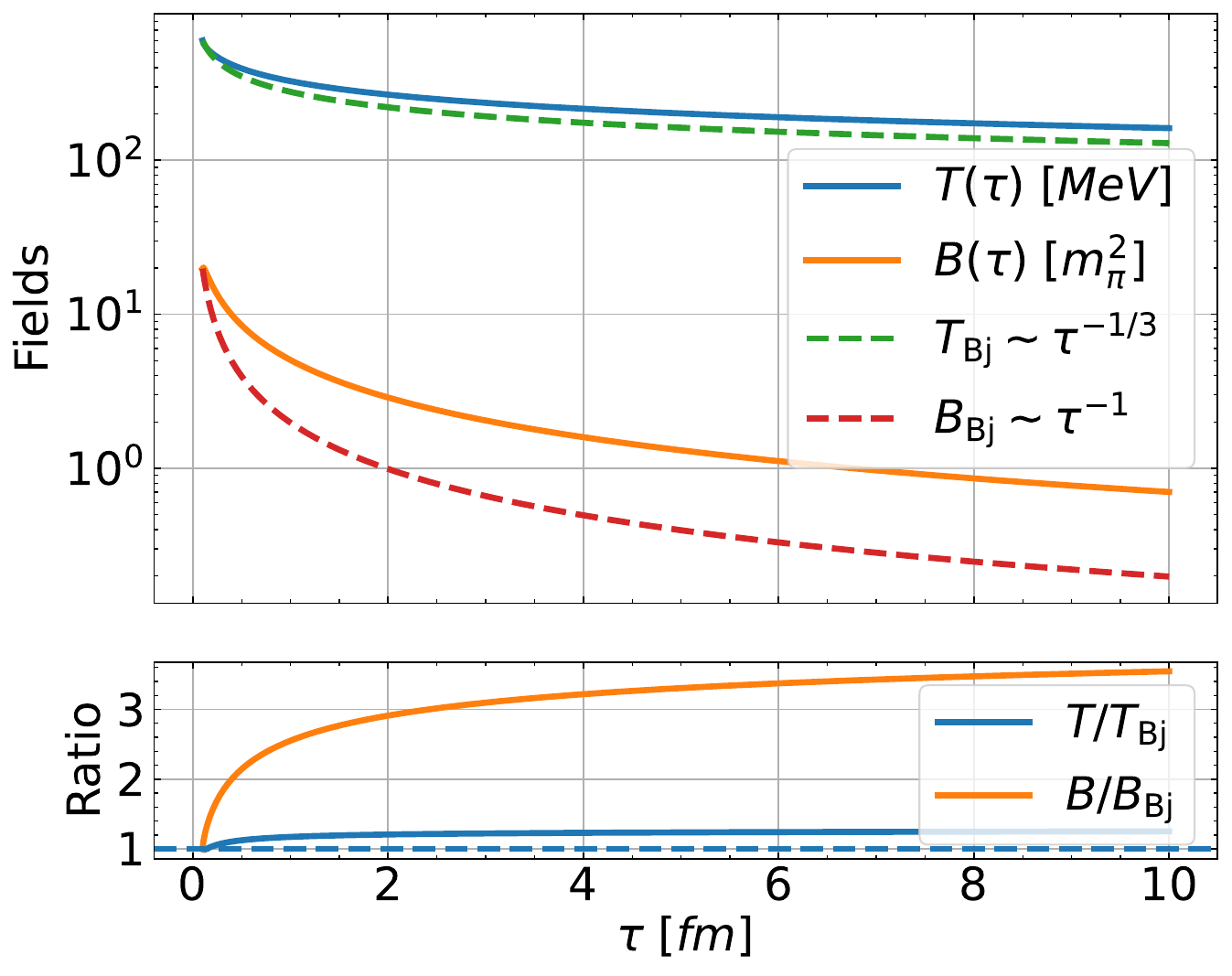}
    \includegraphics[width=\linewidth]{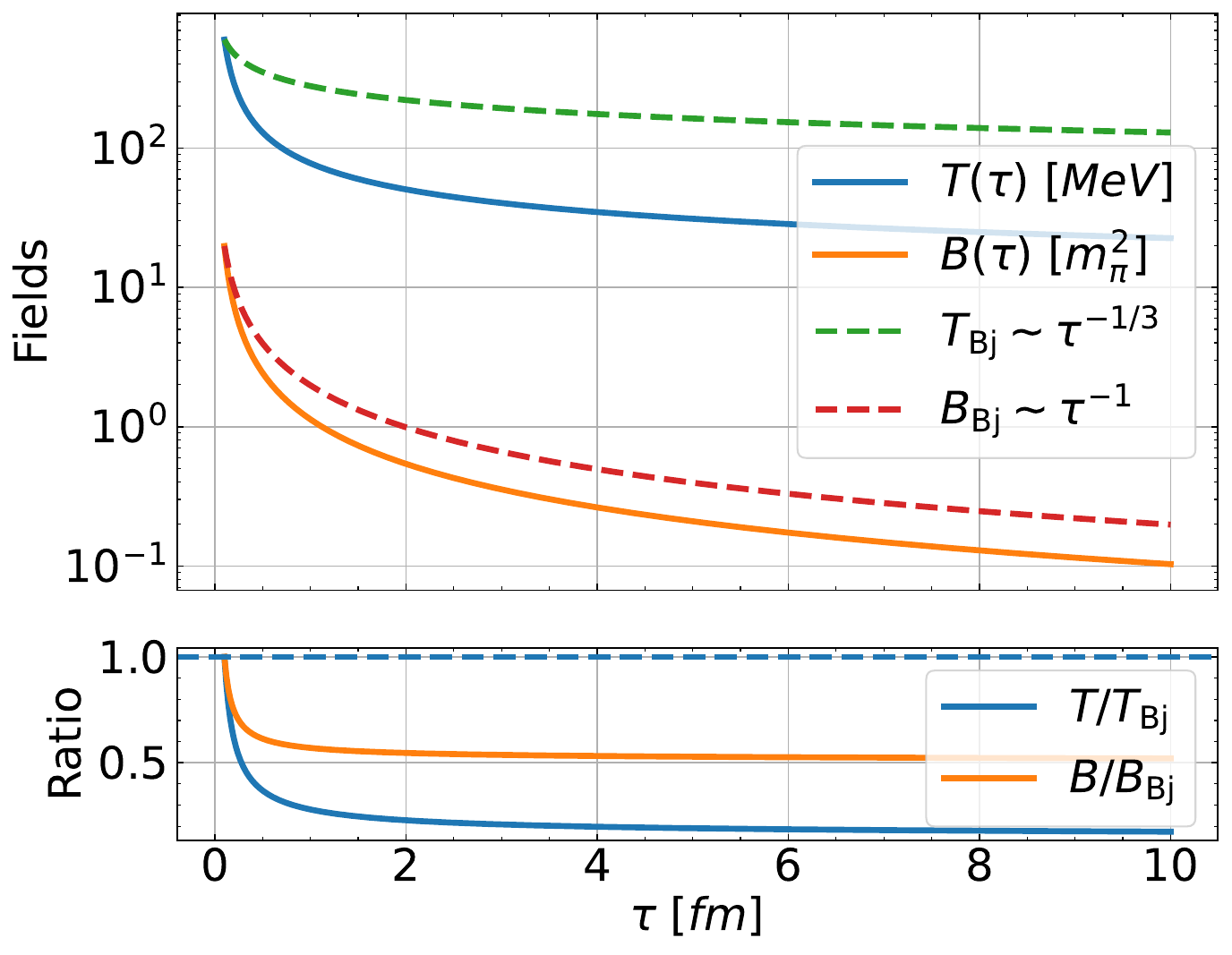}
    \caption{Temperature and magnetic field evolution with all expansion coefficients taken to be non-zero, for initial conditions $T_{\text{in}} = 600~\mathrm{MeV}$ and $B_{\text{in}} = 20\,m_{\pi}^2$. The upper panel corresponds to $(\varepsilon_{01}, \varepsilon_{02}, \varepsilon_{05}) = (0.99,\,0.5,\,0.99)$, $(\pi_{01}, \pi_{02}, \pi_{05}) = (0.7,\,0.5,\,0.99)$, $(\sigma_{01}, \sigma_{02}, \sigma_{05}) = (0.99,\,0.6,\,0.03)$, and $(\beta_{01}, \beta_{02}, \beta_{05}) = (0.6,\,0.4,\,0.1)$. The lower panel shows the corresponding case with all coefficients taken with reversed signs: $(\varepsilon_{01}, \varepsilon_{02}, \varepsilon_{05}) = (-0.99,\,-0.50,\,-0.99)$, $(\pi_{01}, \pi_{02}, \pi_{05}) = (-0.70,\,-0.50,\,-0.99)$, $(\sigma_{01}, \sigma_{02}, \sigma_{05}) = (-0.99,\,-0.60,\,-0.03)$, and $(\beta_{01}, \beta_{02}, \beta_{05}) = (-0.10,\,-0.05,\,-0.03)$.}
    \label{fig:solution_all}
\end{figure}
Although the analysis is performed within a $(0+1)$D boost-invariant framework, this setup captures the essential physical features of fully coupled MHD evolution in an expanding medium.
The resulting dynamics are shown in Fig.~\eqref{fig:solution_all}. The upper panel corresponds to a positive choice of all couplings, while the lower panel shows the case with reversed signs. In our implementation, the coefficients governing the backreaction of the magnetic field on the temperature ($\varepsilon_{05}$, $\pi_{05}$, and $\sigma_{05}$) are taken to be larger in magnitude than the corresponding temperature-to-magnetic-field couplings encoded in $\beta_{\parallel 0i}$. This hierarchy allows us to emphasize the dominant direction of feedback in the system.
We observe that, in the upper panel, the mutual coupling between temperature and magnetic field leads to a simultaneous reduction in their respective decay rates compared to the ideal Bjorken evolution. Notably, the influence of temperature on the magnetic field remains comparatively stronger, even though the corresponding coupling coefficients are smaller, highlighting the nonlinear sensitivity of the magnetic sector to thermal gradients.
In the lower panel, reversing the signs of all coefficients leads to the opposite trend, enhancing the decay rates of both temperature and magnetic field. A clear asymmetry is visible in the ratio plots: in the lower panel, the temperature deviates more significantly from its Bjorken baseline than the magnetic field case. This is because, in the negative-coupling scenario, the $\beta_{\parallel 0i}$ coefficients are taken to be much smaller in magnitude than in the positive case, a choice dictated by numerical stability requirements rather than physical considerations.

\subsection{Impact on Number Density Evolution}
\begin{figure}[h]
    \centering
    \includegraphics[width=\linewidth]{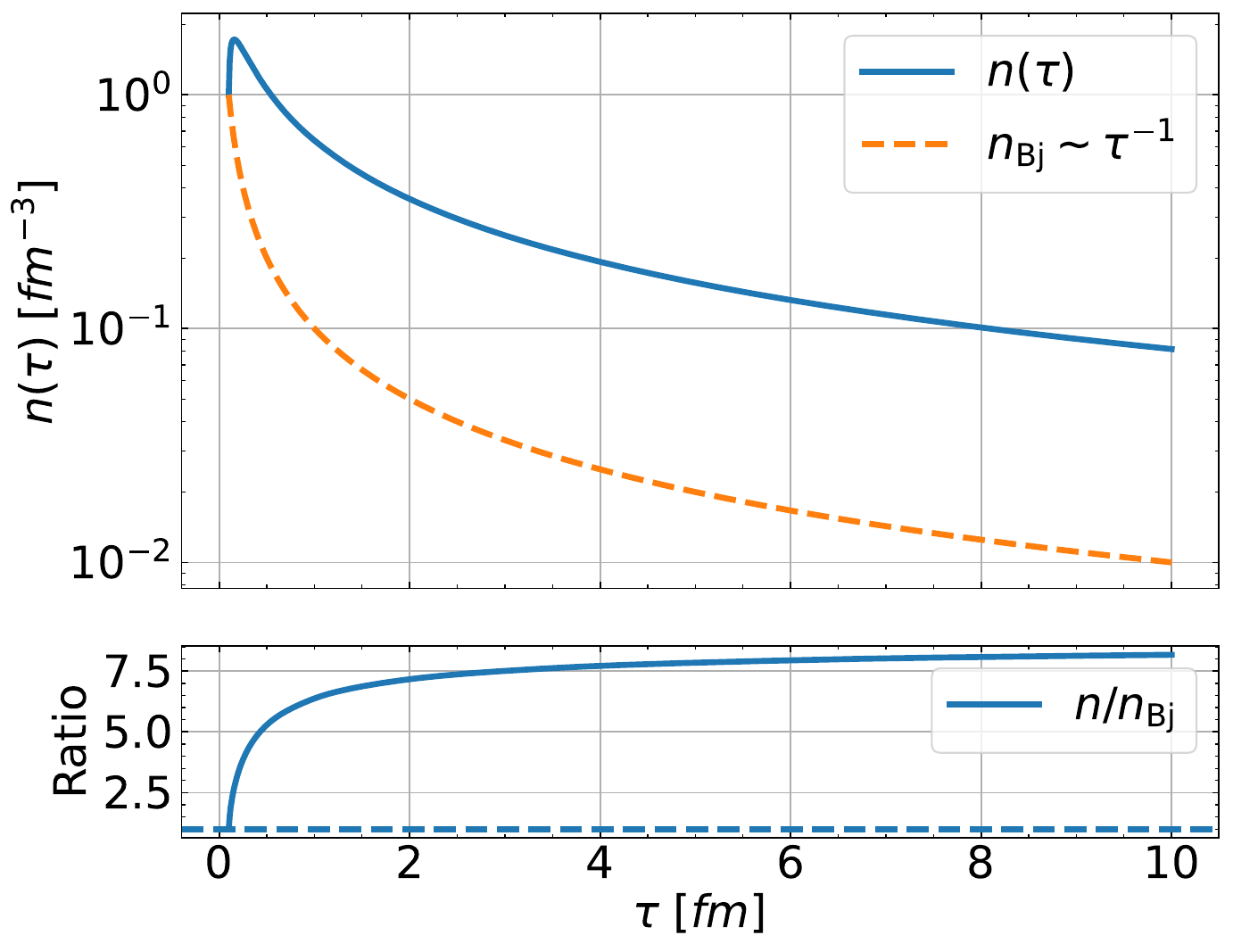}
    \includegraphics[width=\linewidth]{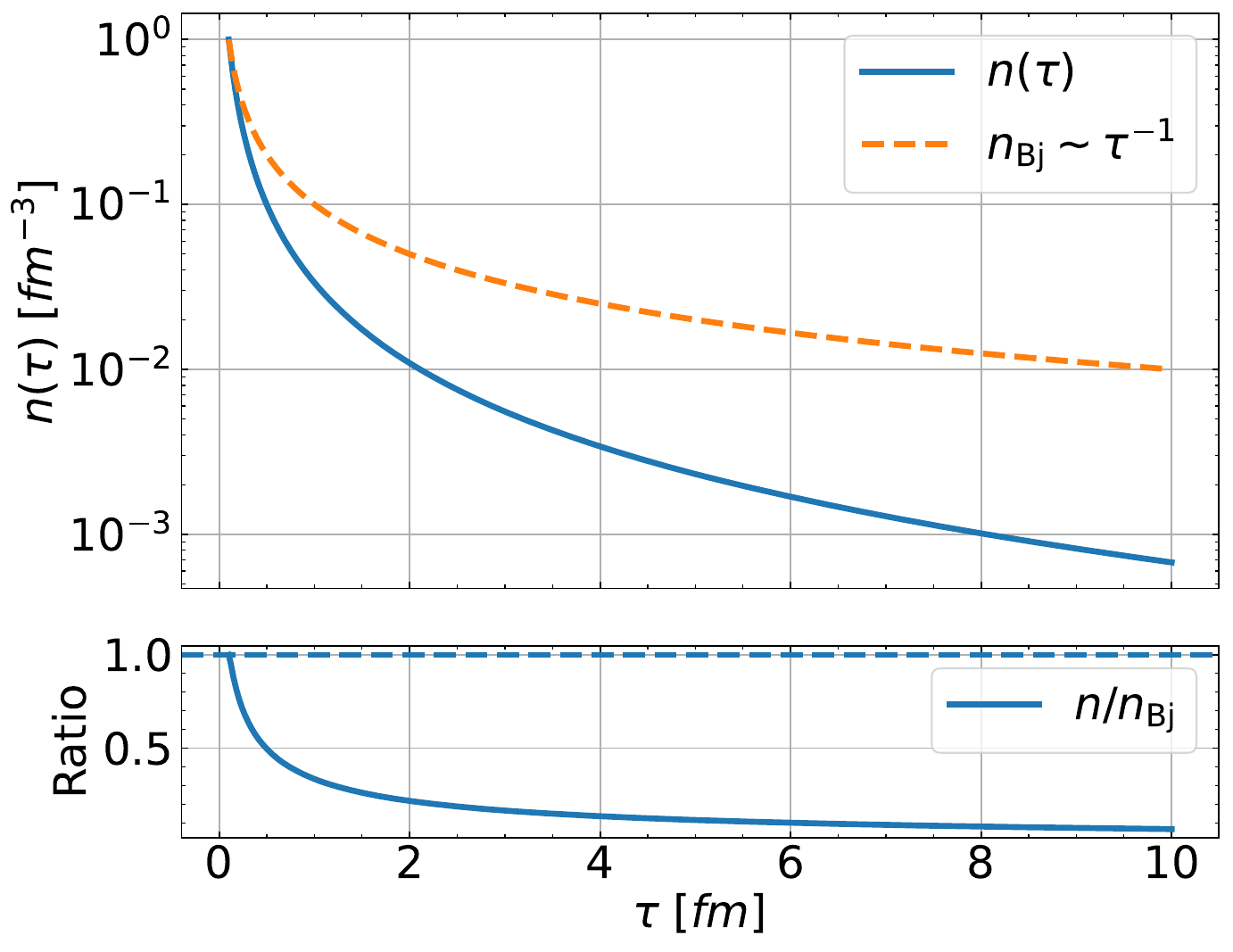}
    \caption{Evolution of the number density with initial condition $n_{\text{in}} = 1~\mathrm{fm}^{-3}$. The upper panel corresponds to $(\nu_{01}, \nu_{02}, \nu_{05}) = (0.1,\,0.03,\,0.01)$, while the lower panel corresponds to $(\nu_{01}, \nu_{02}, \nu_{05}) = (-0.008,\,-0.003,\,-0.007)$. All other expansion coefficients are taken to be positive and non-zero.
}
    \label{fig:numberdensity}
\end{figure}
Finally, having established the coupled evolution of temperature and magnetic field { by activating all the transport parameters } as specified in Fig.~\eqref{fig:solution_all}, we now proceed to implement this setup to study the evolution of the number density.
In addition, we consider a more general set of transport coefficients ($\nu_i$), whose non-zero values introduce an explicit coupling between the magnetic field, temperature, and the number density evolution {equation}. In this framework, the number density dynamics becomes sensitive not only to temperature gradients but also directly to the magnetic field evolution through these couplings.
We illustrate this behaviour in Fig.~\eqref{fig:numberdensity}. The upper panel corresponds to positive values of the coupling coefficients, while the lower panel corresponds to negative values. We observe that positive couplings tend to slow down the decay of the number density, whereas negative couplings enhance its decay rate, effectively reversing the trend.
The stronger enhancement observed in the upper panel, compared to the suppression in the lower panel, arises primarily from the specific choice of coupling strengths used in each case. For both the cases shown, the initial number density is taken to be $1~\mathrm{fm}^{-3}$.
\section{Phenomenological Implications: Dilepton Production}{\label{3}}
Within the fully coupled BDNK--MHD framework, we first determine the self-consistent space-time evolution of the temperature and magnetic field. This evolving background provides the essential input required to connect the theory with experimentally accessible observables.
Among the most penetrating electromagnetic probes in relativistic heavy-ion collisions are {the} dileptons. In the present framework, the dilepton emission rate is evaluated within kinetic theory~\cite{Dwibedi:2025xho}, where it depends explicitly on the properties of the medium and, in particular, on the background magnetic field. The influence of the magnetic field is incorporated through a modification of the microscopic relaxation time, which we parametrize as $\tau_c \rightarrow \tau_c(B)\,.$
\begin{figure}[h]
    \centering
    \includegraphics[width=\linewidth]{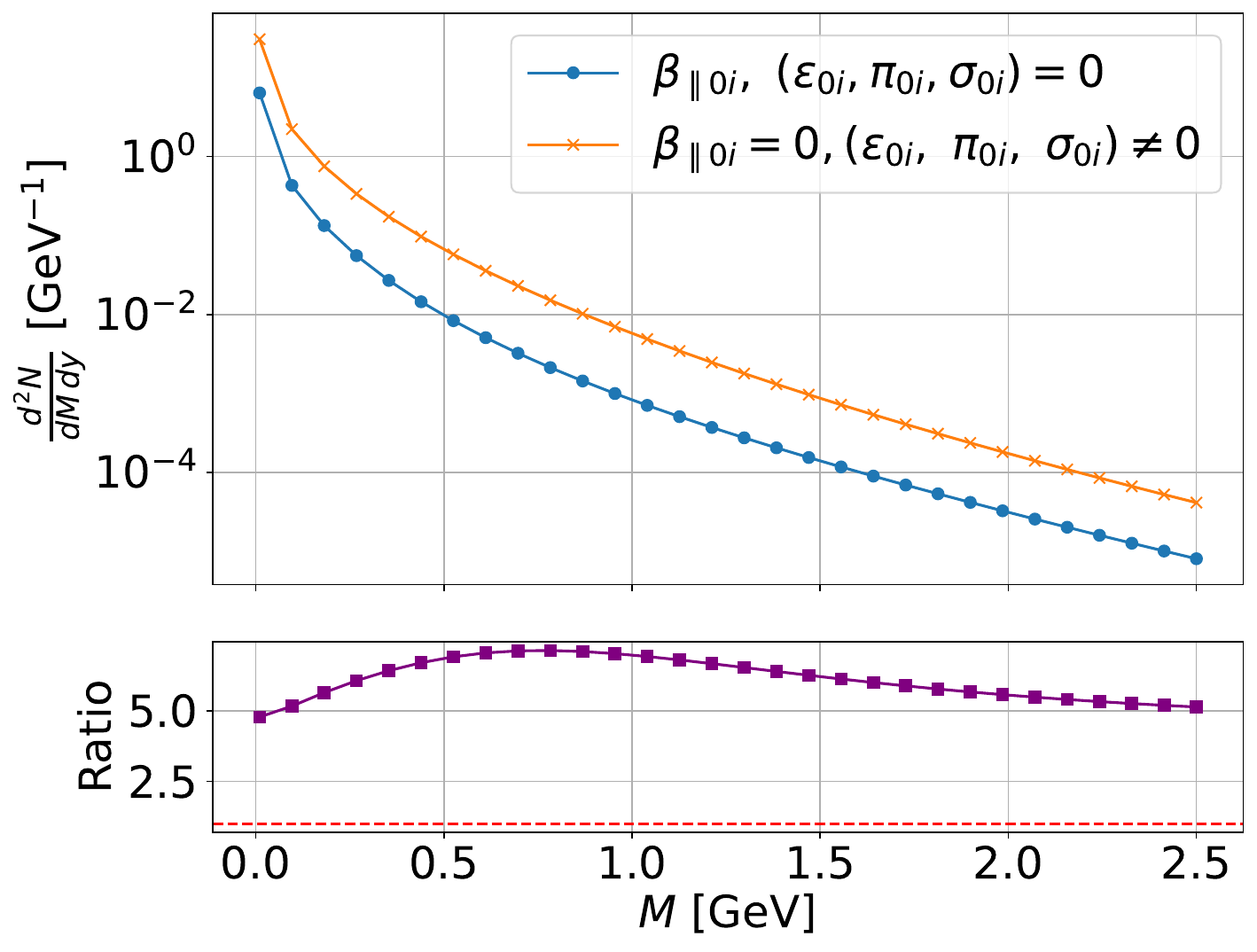}
    \caption{Mass spectra of dileptons at mid-rapidity ($y=0$), integrated over $p_T \in [0.01,\,2]~\mathrm{GeV}$, shown for the case where $\beta_{\parallel 0i}=0$ while all other expansion coefficients are non-zero (orange), compared to the baseline scenario in which all coefficients are set to zero (blue).}
    \label{fig:backreaction_0}
\end{figure}
\begin{figure}
    \centering
    \includegraphics[width=\linewidth]{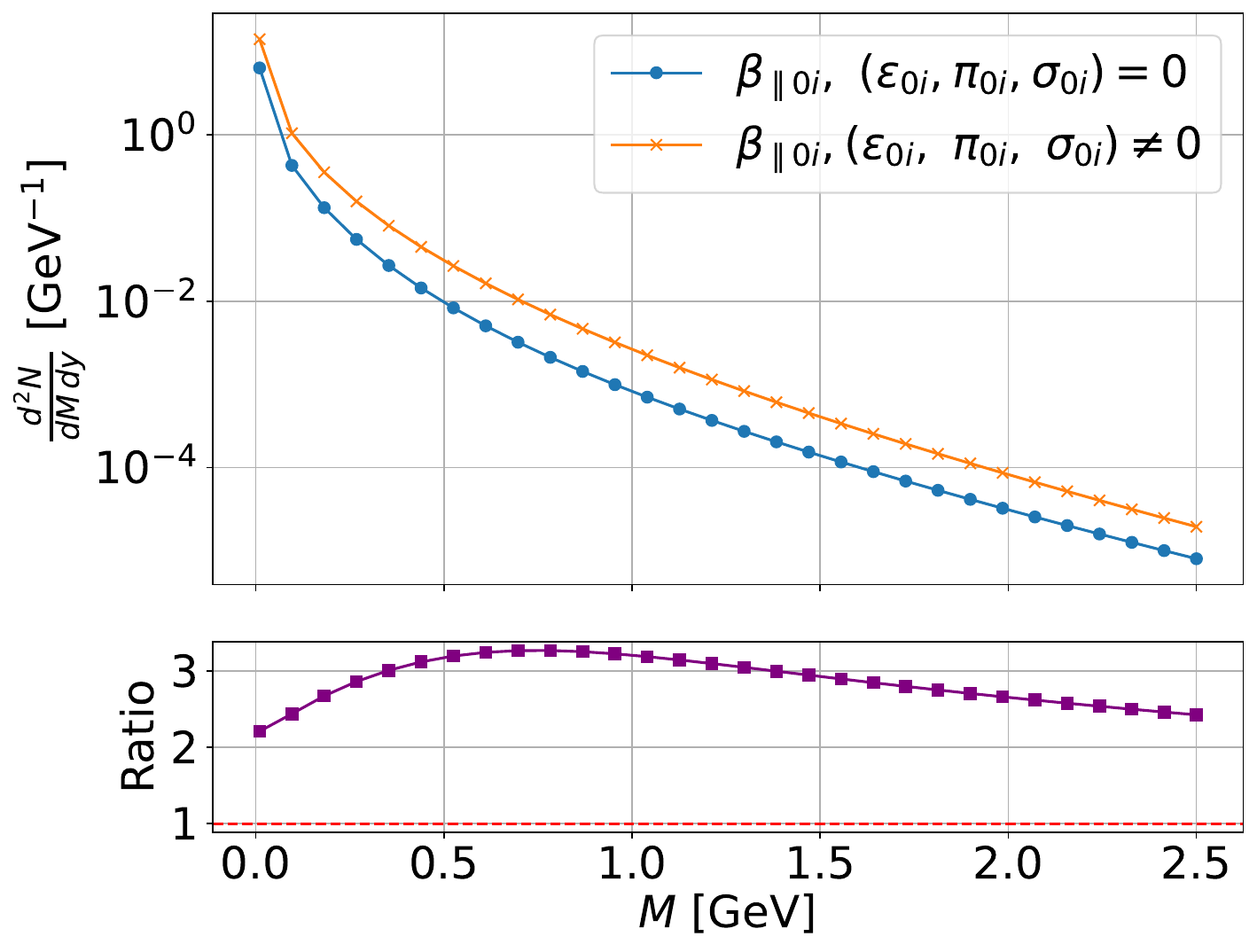}
    \caption{Mass spectra of dileptons at mid-rapidity ($y=0$), integrated over $p_T \in [0.01,\,2]~\mathrm{GeV}$, comparing the case with back-reaction and non-zero expansion coefficients (orange) to the baseline scenario (blue) where all coefficients are set to zero.}
    \label{fig:backreaction_y}
\end{figure}
This dependence reflects the fact that the equilibration dynamics of the medium are altered in the presence of electromagnetic fields. Physically, a stronger magnetic field enhances the interaction rate among charged constituents, leading to a reduction of the relaxation time,
$B \uparrow \;\;\Longrightarrow\;\; \tau_c \downarrow\,$.
As a result, the system approaches local equilibrium more rapidly, and this accelerated equilibration leaves a direct imprint on the dilepton production rate.
There are other works where the magnetic fields are included in the dilepton rate itself using the QFT propagators~\cite{Das:2021fma,Wang:2022jxx,Panda:2025yxw}.
The total dilepton yield is thus obtained by integrating the local emission rate over the full space-time evolution of the system,
\begin{eqnarray}
\frac{dN}{d^{4}q}
= \int d^{4}x \, \frac{d^{4}N}{d^{4}x \, d^{4}q}
= \int \tau \, d\tau \, dx \, dy \, d\eta \;
\frac{d^{4}N}{d^{4}x \, d^{4}q},\,
\end{eqnarray}
where \( d^{4}x = \tau \, d\tau \, dx \, dy \, d\eta \) denotes the invariant volume element in Milne coordinates. The corresponding momentum space differential measure is given by $d^{4}q = q_T \, dq_T \, dM^2 \, dy \, d\phi_p\,,$
with \( q_T \), \( M \), \( y \), and \( \phi_p \) representing the transverse momentum, invariant mass, rapidity, and azimuthal angle of the dilepton pair produced, respectively.
In this work, we adopt the Bjorken flow profile for the fluid velocity and assume transverse homogeneity, consistent with the underlying symmetries of the system. The space–time integration is therefore carried out over the ranges \(x,y \in [-25,\,25]~\mathrm{fm}\), \(\eta \in [-3,\,3]\), and \(\tau \in [0.01,\,10]~\mathrm{fm}\).
Finally, to obtain the invariant mass spectra, we integrate over the momentum variables, taking \( \phi_p \in [0,2\pi] \) and \( q_T \in [0.01,\,2]~\mathrm{GeV} \), and restrict our analysis to mid-rapidity (\(y=0\)).
Here, we mainly integrate over the fluid cells whose local temperature is over $130~\text{MeV}$.
Within this numerical setup, we investigate the physical effect of backreaction, namely the influence of the magnetic field evolution on the temperature dynamics. This is systematically explored by selectively switching the coefficients \( \beta_{\parallel i} \) on and off.
In Fig.~\eqref{fig:backreaction_0}, we consider the case where $\beta_{\parallel 0i} = 0\,, \quad
\varepsilon_{0i}, \ \pi_{0i}, \ \sigma_{0i} \neq 0\,,$
 this eliminates the effect of thermal gradients on the magnetic field evolution. We compare the resulting dilepton mass spectra with the baseline scenario in which all transport coefficients are set to zero. In this configuration, we observe a pronounced enhancement in the intermediate mass region (\(0.3\text{--}1.2~\mathrm{GeV}\)), reaching up to a factor of \(\sim 5\). This behavior indicates that the inclusion of expansion coefficients effectively slows down the cooling of the system, leading to a sustained higher temperature and consequently an enhanced dilepton yield. This trend is clearly reflected in the ratio plot shown in Fig.~\eqref{fig:backreaction_0}.
In contrast, Fig.~\eqref{fig:backreaction_y} presents the scenario where the same setup is considered, but with the coefficients \( \beta_{\parallel i} \) switched on. In this case, the enhancement relative to the ideal baseline is reduced from approximately a factor of \(5\) to about \(3\). This reduction can be understood as a consequence of the coupling between temperature gradients and magnetic field evolution. With nonzero \( \beta_{\para i} \) {parameters}, the magnetic field dynamically responds to the evolving temperature, which in turn feeds back into the system and accelerates the overall cooling rate. As a result, the dilepton yield is correspondingly suppressed compared to the previous case.

\section{Conclusion}{\label{4}}

In this work, we have investigated the coupled evolution of temperature, magnetic field, and number density within a boost-invariant $(0+1)$D framework based on the BDNK {type} formulation of relativistic MHD. By systematically varying the transport {parameters} associated with the {first-order} gradient expansion, we were able to disentangle and isolate the distinct physical mechanisms governing the mutual interplay between thermal and electromagnetic sectors.
We first established a baseline scenario in which all {non-equilibrium correction, i.e., all transport parameters } are set to zero. In this limit, the system reproduces the expected {ideal} Bjorken scaling behavior, confirming the consistency of our numerical implementation. Although the magnetic field contributes to the energy conservation equation through the competing $B\,\dot{B}$ and $B^2$ terms, their near cancellation mainly for the Bjorken scaling case renders the net effect subleading, resulting in an almost ideal hydrodynamic evolution.
Turning on the $\beta_{\parallel i}$ coefficients, we examined the influence of temperature gradients on the magnetic field evolution. We find that this coupling can significantly modify the decay rate of the magnetic field, with positive (negative) coefficients leading to a slower (faster) decay. This modified magnetic field evolution feeds back into the temperature dynamics, where the dissipative contribution associated with $B^2$ term typically dominates over the $B\,\dot{B}$ term. As a result, the temperature evolution exhibits a comparatively milder but non-negligible deviation from the Bjorken baseline, that is the decay of temperature gets faster.
Conversely, by activating the coefficients $\varepsilon_{05}$, $\pi_{05}$, and $\sigma_{05}$, we isolated the effect of magnetic fields on the thermal evolution. In this case, we observe that the temperature decay can either be slowed down or accelerated depending on the sign of the couplings, reflecting a competition between temperature-gradient and magnetic field–gradient contributions. Importantly, this feedback is quantitatively weaker than the inverse effect of temperature on the magnetic field, indicating an intrinsic asymmetry in the coupling structure of the system.
When all transport coefficients are switched on simultaneously, the system exhibits fully coupled nonlinear dynamics. In this regime, both temperature and magnetic field evolution deviate significantly from the ideal Bjorken scaling. The results reveal that the magnetic sector is particularly sensitive to thermal gradients, even when the corresponding coupling coefficients are smaller, highlighting the nonlinear nature of the underlying dynamics. The interplay between different transport channels ultimately determines whether the system experiences an effective delay or enhancement of cooling.
We further extended our analysis to the evolution of the number density by incorporating the $\nu_i$ coefficients. The number density is found to inherit sensitivity to both temperature and magnetic field evolution through these couplings. Positive (negative) values of $\nu_i$ lead to a slower (faster) decay of the number density, demonstrating that dissipative and electromagnetic effects can substantially modify conserved charge dynamics in an expanding medium.

Finally, we explored the phenomenological implications of our results for dilepton production. By incorporating the magnetic-field dependence of the relaxation time, $\tau_c \to \tau_c(B)$, we established a direct connection between the evolving electromagnetic background and experimentally observable dilepton yields. We find that the inclusion of {transport parameters} can significantly enhance the dilepton yield in the intermediate mass region, primarily due to a delayed cooling of the system. However, once the backreaction of temperature gradients on the magnetic field is included through nonzero $\beta_{\parallel i}$ coefficients, this enhancement is reduced, reflecting a faster effective cooling driven by the coupled dynamics.
Overall, our study demonstrates that the interplay between temperature, magnetic field, and dissipative transport coefficients plays a crucial role in determining the space-time evolution of the medium created in relativistic heavy-ion collisions. These effects leave measurable imprints on electromagnetic observables such as dilepton spectra, emphasizing the importance of a fully coupled and self-consistent MHD description. Future extensions to higher-dimensional simulations and more realistic initial conditions will be essential to further quantify these effects and connect them directly with experimental data.

\begin{acknowledgments}
We would like to thank Arpan Das for useful discussions. Rajesh Biswas is supported, in part, by the Polish National Science Centre (NCN) Sonata Bis Grant 2019/34/E/ST3/00405. Ankit Kumar Panda is supported by Central China Normal University and by NSFC Grants No.12435009.
\end{acknowledgments}

\bibliographystyle{apsrev4-2}
\bibliography{biblio}

\end{document}